\documentclass[
    ,final            % use final for the camera ready runs
  ]
  {aipproc}

\layoutstyle{6x9}
\usepackage{graphicx}

\begin{document}

\title{Excited-State Hadrons using the Stochastic LapH Method}


\author{C.~Morningstar}{
  address={Dept.~of Physics, Carnegie Mellon University, 
           Pittsburgh, PA 15213, USA}
}

\author{A.~Bell}{
  address={Dept.~of Physics, Carnegie Mellon University, 
           Pittsburgh, PA 15213, USA}
}

\author{J.~Bulava}{
  address={NIC, DESY, Platanenallee 6, D-15738, Zeuthen, Germany}
}

\author{J.~Foley}{
  address={Physics Department, University of Utah, 
           Salt Lake City, UT 84112, USA}
}

\author{K.J.~Juge}{
  address={Dept.~of Physics, University of the Pacific, Stockton, CA 95211, USA}
}

\author{D.~Lenkner}{
  address={Dept.~of Physics, Carnegie Mellon University, 
           Pittsburgh, PA 15213, USA}
}

\author{C.H.~Wong}{
  address={Dept.~of Physics, Carnegie Mellon University, 
           Pittsburgh, PA 15213, USA}
}

\begin{abstract}
 Progress in computing the spectrum of excited baryons and mesons in lattice
 QCD is described.  Large sets of spatially-extended hadron operators are
 used. A new method of stochastically estimating the low-lying effects of 
 quark propagation is utilized which allows reliable
 determinations of temporal correlations of both single-hadron and multi-hadron 
 operators.  The method is tested on the $\eta,\sigma,\omega$ mesons.
\end{abstract}

\maketitle

We are currently carrying out computations of the excitation spectrum of QCD
in finite volume with \textit{ab initio} Markov-chain Monte Carlo path 
integrations on anisotropic space-time lattices.  Our first results using
two flavors of dynamical quarks were reported in Ref.~\cite{spectrum2009},
and our most recent results can be found in Ref.~\cite{wallace2010}.
Such calculations 
are very challenging.  Computational limitations cause simulations
to be done with quark masses that are unphysically large, leading to pion
masses that are especially heavier than observed.  The use of carefully
designed quantum field operators is crucial for accurate determinations of
low-lying energies. To study a particular state of interest, the energies of
all states lying below that state must first be extracted, and 
as the pion gets lighter in lattice QCD simulations, more and more multi-hadron 
states lie below the masses of the excited resonances.  The evaluation
of correlations involving multi-hadron operators contains new challenges since
not only must initial to final time quark propagation be included, but also 
final to final time quark propagation must be incorporated. 

The use of operators whose correlation functions $C(t)$ attain their
asymptotic form as quickly as possible is crucial for reliably
extracting excited hadron masses.  An important ingredient in constructing
such hadron operators is the use of smeared fields.  Operators constructed
from smeared fields have dramatically reduced mixings with the high frequency
modes of the theory.  Both link-smearing\cite{stout} and quark-field 
smearing\cite{distillation2009} must
be applied.  Since excited hadrons are expected to be large objects, 
the use of spatially extended operators is another key ingredient in
the operator design and implementation.  A more detailed discussion
of these issues can be found in Ref.~\cite{baryons2005A}.

\begin{figure}
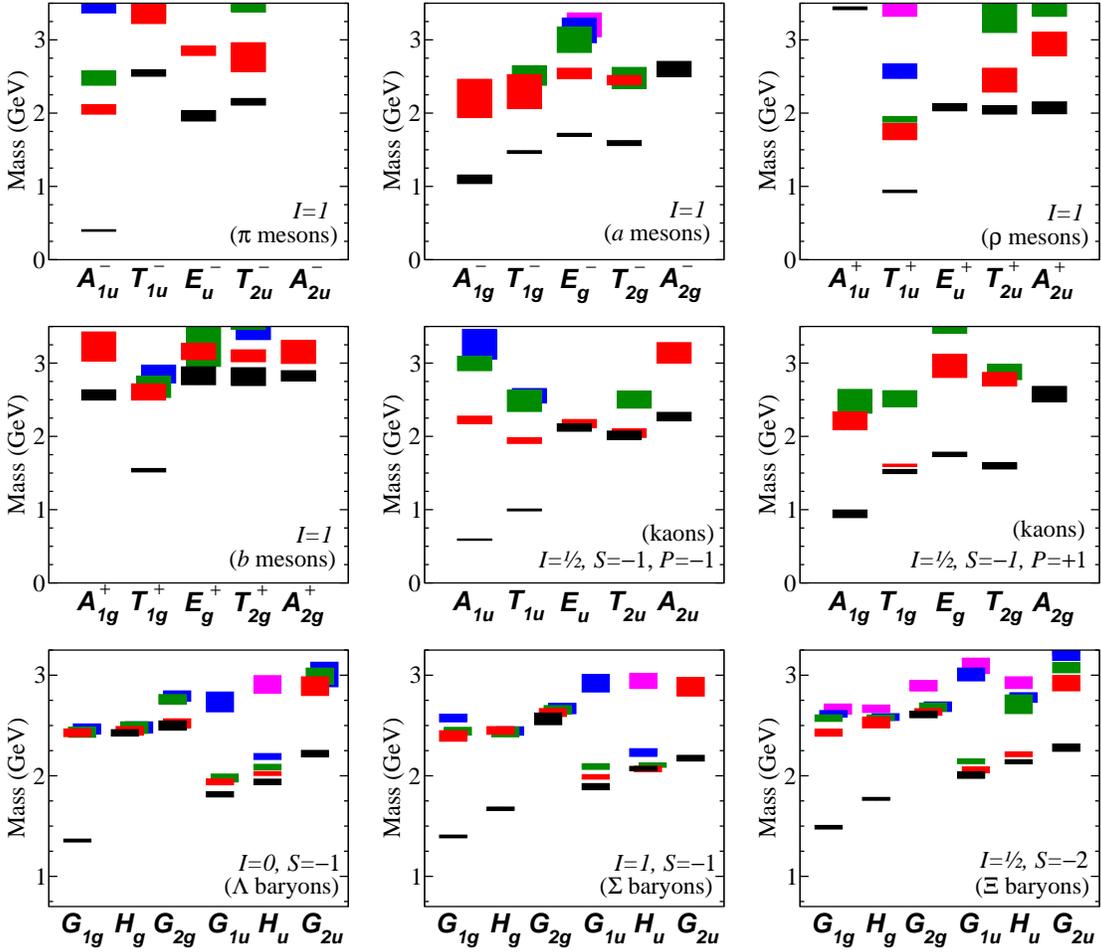

\begin{minipage}{5.9in}
\hspace*{0mm}
\includegraphics[width=1.8in,bb=31 34 523 461]{pi_mesons_pruning.eps}\quad
\includegraphics[width=1.8in,bb=31 34 523 461]{a_mesons_pruning.eps}\quad
\includegraphics[width=1.8in,bb=31 34 523 461]{rho_mesons_pruning.eps}\\[3mm]
\hspace*{0mm}
\includegraphics[width=1.8in,bb=31 34 523 461]{b_mesons_pruning.eps}\quad
\includegraphics[width=1.8in,bb=31 34 523 461]{kaons_oddp_pruning.eps}\quad
\includegraphics[width=1.8in,bb=31 34 523 461]{kaons_evenp_pruning.eps}\\[3mm]
\hspace*{0mm}
\includegraphics[width=1.8in,bb=31 34 523 461]{lambda_baryons_pruning.eps}\quad
\includegraphics[width=1.8in,bb=31 34 523 461]{sigma_baryons_pruning.eps}\quad
\includegraphics[width=1.8in,bb=31 34 523 461]{xi_baryons_pruning.eps}
\end{minipage}
\caption{
Hadron operator selection: low-statistics simulations have been performed
to study the hundreds of single-hadron operators produced by our group-theoretical
construction.  A ``pruning" procedure was followed in each channel to select
good sets of between six to a dozen operators.  The plots above show the
stationary-state energies extracted to date from correlation matrices of the finally
selected single-hadron operators.  Results were obtained using between 50 to 100 
configurations on a $16^3\times 128$ anisotropic lattice for $N_f=2+1$ quark flavors
with spacing $a_s\sim 0.12$~fm, $a_s/a_t\sim 3.5$, and quark masses such that
$m_\pi\sim 380$~MeV.  Each box indicates the energy of one stationary 
state; the vertical height of each box indicates the statistical error.
\label{fig:pruning}}
\end{figure}

A large effort was undertaken during the last two years to select optimal
sets of baryon and meson operators in a large variety of isospin sectors
and for zero momentum and non-zero on-axis, planar-diagonal, and cubic-diagonal
momenta. Low-statistics Monte Carlo computations were done to accomplish these
operator selections using between 50 to 100 configurations on a $16^3\times 128$ 
anisotropic lattice for $N_f=2+1$ quark flavors with spacing $a_s\sim 0.12$~fm, 
$a_s/a_t\sim 3.5$, and quark masses such that the pion has mass around 380~MeV. 
Stationary-state energies using the finally selected operator sets are shown in
Fig.~\ref{fig:pruning}.  The nucleon, $\Delta$, $\Xi$, $\Sigma$, and $\Lambda$
baryons were studied, and light isovector and kaon mesons were investigated. 
Hundreds of operators were studied, and optimal sets containing eight or so
operators in each symmetry channel were found.  Future computations will focus
solely on the operators in the optimal sets. 

A comprehensive picture of resonances requires that we go beyond a
knowledge of the ground state mass in each symmetry channel and obtain 
the masses of the lowest few states in each channel.  This necessitates
the use of \textit{matrices} of correlation functions.  
Rather than evaluating a single correlator $C(t)$, we determine
a matrix of correlators
$
C_{ij}(t) = \langle  O_i(t) O^{\dagger}_j
(t_0) \rangle \label{eq:corrs_cons},
$
where $\{  O_i; i = 1,\dots,N \}$ are a basis of interpolating
operators with given quantum numbers.  We then solve the generalized
eigenvalue equation
$
C(t) u = \lambda(t, t_0) C(t_0) u
$
to obtain a set of real (ordered) eigenvalues $\lambda_n(t,t_0)$,
where $\lambda_0 \ge \lambda_1 \ge\dots \ge \lambda_{N-1}$.  At large 
Euclidean times,
these eigenvalues then delineate between the different masses
$
\lambda_n (t,t_0) \longrightarrow e^{ -M_n (t-t_0)} 
 + O (e^{ - \Delta M_n (t-t_0)}),
$
where $\Delta M_n = \mbox{min} \{ \mid M_n - M_i \mid : i \ne n \}$.
The eigenvectors $u$ are orthogonal with metric $C(t_0)$, and
the eigenvectors yield information about the
structure of the states.  

\begin{figure}
\includegraphics[width=3.5in,bb=0 30 743 324]{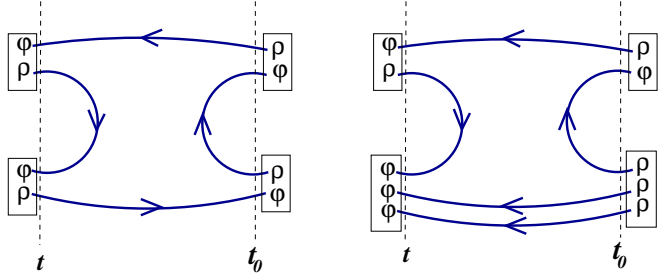}
\caption{
Diagrams of multi-hadron correlators that require having $\varrho$ 
noise sources on the later time $t$. Solution vectors are denoted
by $\varphi$. (Left) A two-meson correlator. (Right) The correlator
of a baryon-meson system.
\label{fig:multicorr}}
\end{figure}

To study a particular eigenstate of interest with this method, all 
eigenstates lying below that state must first be extracted, and as the pion gets
lighter in lattice QCD simulations, more and more multi-hadron states will
lie below the excited resonances.  A \textit{good} baryon-meson operator of total
zero momentum is typically a superposition of local interpolating fields
at all sites on a time slice of the lattice.  In the evaluation of 
the temporal correlations of such a multi-hadron operator, it is not possible to
completely remove all summations over the spatial sites on the source time-slice using
translation invariance.  Hence, the need for estimates of the quark propagators from 
all spatial sites on a time slice to all spatial sites on another time slice
cannot be sidestepped.  Some correlators will involve diagrams with
quark lines originating at the sink time $t$ and terminating at the
same sink time $t$ (see Fig.~\ref{fig:multicorr}), so quark propagators involving
a large number of starting times $t$ must also be handled.

Finding better ways to stochastically estimate slice-to-slice quark propagators
is crucial to the success of our excited-state hadron spectrum project
at lighter pion masses.  We have developed and tested a new scheme 
which combines a new way of smearing the quark field with a new way of 
introducing noise. The new quark-field smearing scheme, called 
Laplacian Heaviside (LapH), has been described in Ref.~\cite{distillation2009}
and is defined by
\begin{equation}
\widetilde{\psi}(x) = 
 \Theta\left(\sigma_s^2+\widetilde{\Delta}\right)\psi(x),
\end{equation}
where $\widetilde{\Delta}$ is the three-dimensional covariant Laplacian
in terms of the stout-smeared gauge field and $\sigma_s$ is the smearing
parameter.  The gauge-covariant Laplacian operator is
ideal for smearing the quark field since it is one of the simplest operators
that locally averages the field in such a way that all relevant symmetry
transformation properties of the original field are preserved.  
Let $V_\Delta$ denote the unitary matrix whose columns are the eigenvectors
of $\widetilde{\Delta}$, and let $\Lambda_\Delta$ denote a diagonal matrix
whose elements are the eigenvalues of $\widetilde{\Delta}$ such that
$
    \widetilde{\Delta}=V_\Delta\ \Lambda_\Delta\ V_\Delta^\dagger.
$
The LapH smearing matrix is then given by
$
    S = V_\Delta\ \Theta\left(\sigma_s^2+\Lambda_\Delta\right)
   \ V_\Delta^\dagger. 
$
Let $V_s$ denote the matrix whose columns are in one-to-one correspondence
with the eigenvectors associated with the $N_v$ lowest-lying eigenvalues of 
$-\widetilde{\Delta}$ on each time slice.  Then our LapH smearing matrix 
is well approximated by the Hermitian matrix
$
   S=V_s\ V_s^\dagger.
$
Evaluating the temporal correlations of our hadron operators
requires combining Dirac matrix elements associated with various quark 
lines $Q$.  Since we construct our hadron operators out of
covariantly-displaced, smeared quark fields, each and every quark line 
involves the following product of matrices:
\begin{equation}
 Q = D^{(j)}SM^{-1}SD^{(k)\dagger},
\end{equation}
where $D^{(i)}$ is a gauge-covariant displacement of type $i$. 
An exact treatment of such a quark line is very costly, so we resort
to stochastic estimation.

Random noise vectors $\eta$ whose expectations satisfy
$E(\eta_i)=0$ and $E(\eta_i\eta_j^\ast)=\delta_{ij}$ are useful for 
stochastically estimating the inverse of a large matrix $M$ 
as follows\cite{alltoall}.  Assume that for each of $N_R$ 
noise vectors, we can solve the following
linear system of equations: $M X^{(r)}=\eta^{(r)}$ for $X^{(r)}$.
Then $X^{(r)}=M^{-1}\eta^{(r)}$, and $E( X_i \eta_j^\ast ) = M^{-1}_{ij}$
so that a Monte Carlo estimate of $M_{ij}^{-1}$ is given by
$
  M_{ij}^{-1} \approx \lim_{N_R\rightarrow\infty}\frac{1}{N_R}
 \sum_{r=1}^{N_R} X_i^{(r)}\eta_j^{(r)\ast}.
$
Unfortunately, this equation usually produces stochastic estimates with 
variances which are much too large to be useful.  Variance reduction is
done by \textit{diluting} the noise vectors.
A given dilution scheme can be viewed as the application of a complete
set of projection operators $P^{(a)}$. Define
$
  \eta^{[a]}_k=P^{(a)}_{kk^\prime}\eta_{k^\prime} ,
$
and further define $X^{[a]}$ as the solution of
$
   M_{ik}X^{[a]}_k=\eta^{[a]}_i,
$
then we have
\begin{equation}
   M_{ij}^{-1}\approx \lim_{N_R\rightarrow\infty}\frac{1}{N_R}
 \sum_{r=1}^{N_R} \sum_a X^{(r)[a]}_i\eta^{(r)[a]\ast}_j.
\label{eq:diluted}
\end{equation}
The use of $Z_4$ noise ensures
zero variance in the diagonal elements $E(\eta_i\eta_i^\ast)$.

The effectiveness of the variance reduction depends on the
projectors chosen.  With LapH smearing, noise vectors 
$\rho$ can be introduced \textit{only in the LapH subspace}.  The noise
vectors $\rho$ now have spin, time, and Laplacian eigenmode number
as their indices.  Color and space indices get replaced by Laplacian
eigenmode number.  Again, each component of $\rho$ is a random
$Z_4$ variable so that $E(\rho)=0$ and $E(\rho\rho^\dagger)=I_d$.
Dilution projectors $P^{(b)}$ are now matrices in the LapH subspace. 
In the stochastic LapH method, a quark line on a gauge configuration
is estimated using
\begin{equation}
 Q_{uv} \approx \frac{1}{N_R}\sum_{r=1}^{N_R}\sum_b  \varphi^{(r)[b](j)}_u
  \  \varrho^{(r)[b](k)\ast}_v,
\end{equation}
where the subscripts $u,v$ are compound indices combining space, time, color,
and spin, and for a noise vector labelled by index $r$, 
displaced-smeared-diluted quark source 
and quark sink vectors can be defined by
\begin{eqnarray}
 \varrho^{(r)[b](j)} &=&  D^{(j)} V_s P^{(b)}\rho^{(r)},\\
 \varphi^{(r)[b](j)} &=& D^{(j)} S M^{-1}\ V_s P^{(b)}\rho^{(r)}. 
\end{eqnarray}

Our dilution projectors are
products of time dilution, spin dilution, and Laph eigenvector dilution
projectors.  For each type (time, spin, Laph eigenvector) of dilution, we
studied four different dilution schemes.  Let $N$ denote the dimension
of the space of the dilution type of interest.  For time dilution, $N=N_t$
is the number of time slices on the lattice.  For spin dilution, $N=4$ is
the number of Dirac spin components.  For Laph eigenvector dilution, $N=N_v$
is the number of eigenvectors retained.  The four schemes we studied
are defined below:
\[ \begin{array}{lll}
P^{(a)}_{ij} = \delta_{ij},              & a=0,& \mbox{(no dilution)} \\
P^{(a)}_{ij} = \delta_{ij}\ \delta_{ai},  & a=0,\dots,N-1 & \mbox{(full dilution)}\\
P^{(a)}_{ij} = \delta_{ij}\ \delta_{a,\, \lfloor Ki/N\rfloor}& a=0,\dots,K-1, & \mbox{(block-$K$)}\\
P^{(a)}_{ij} = \delta_{ij}\ \delta_{a,\, i\bmod K} & a=0,\dots,K-1, & \mbox{(interlace-$K$)}
\end{array}\]
where $i,j=0,\dots,N-1$, and we assume $N/K$ is an integer.  We use a triplet
(T, S, L) to specify a given dilution scheme, where ``T" denote time,
``S" denotes spin, and ``L" denotes Laph eigenvector dilution.  The schemes
are denoted by 1 for no dilution, F for full dilution, and B$K$ and I$K$ for
block-$K$ and interlace-$K$, respectively.  For example, full time and spin
dilution with interlace-8 Laph eigenvector dilution is denoted by
(TF, SF, LI8).  Introducing diluted noise in this
way produces correlation functions with significantly reduced variances,
yielding nearly an order of magnitude reduction in the statistical error
over previous methods.  The volume dependence of this new method was found 
to be very mild, allowing the method to be useful on large lattices.
For all forward-time quark lines, we use dilution scheme (TF, SF, LI8),
and for all same-sink-time quark lines, we use (TI16, SF, LI8).

\begin{figure}
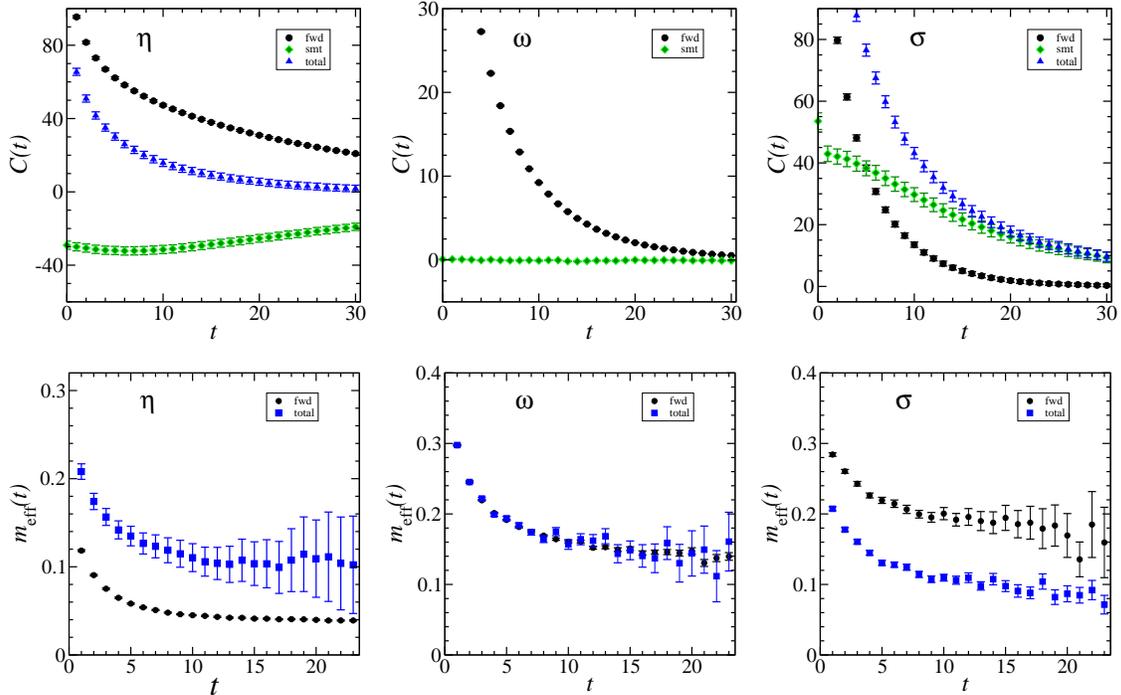

\begin{minipage}{5.9in}
\includegraphics[width=1.8in,bb=21 34 523 530]{corr860.PS.eps}\quad
\includegraphics[width=1.8in,bb=21 34 523 530]{corr860.V.eps}\quad
\includegraphics[width=1.8in,bb=21 34 523 530]{corr860.S.eps}\\[3mm]
\includegraphics[width=1.8in,bb=21 34 543 530]{effmass860.PS.eps}\quad
\includegraphics[width=1.8in,bb=21 34 543 530]{effmass860.V.eps}\quad
\includegraphics[width=1.8in,bb=21 34 543 530]{effmass860.S.eps}\\[-7mm]
\end{minipage}
\caption{
Correlators $C(t)$ and effective masses $m_{\rm eff}(t)$ against
temporal separation $t$ for single-site operators which produce
the isoscalar pseudoscalar $\eta$, vector $\omega$, and scalar $\sigma$ mesons.
Results were obtained using 198 configurations with $N_f=2+1$ flavors of
quark loops on a $24^3\times 128$ anisotropic lattice with spacing
$a_s\sim 0.12$~fm and aspect ratio $a_s/a_t\sim 3.5$ for a pion mass
$m_\pi\sim 220$~MeV. In the legends, ``fwd" refers to contributions
from the diagram containing only forward-time source-to-sink quark lines,
``smt" refers to contributions from the diagram containing only
quark lines that originate and terminate at the same time.  For the 
$\sigma$ channel, the ``smt" contribution has a vacuum expectation value
subtraction. Forward-time quark lines use dilution scheme (TF, SF, LI8)
and same-time quark lines use (TI16, SF, LI8).
\label{fig:isoscalars860}}
\end{figure}

Results for three isoscalar mesons are shown in Fig.~\ref{fig:isoscalars860}.
Such mesons are notoriously difficult to study in lattice QCD, but the new
method appears to produce estimates of their temporal correlations with 
unprecedented accuracy.  These plots suggest that evaluating 
correlation functions involving our multi-hadron operators will be feasible 
with the stochastic LapH method.

We are currently carrying out these spectrum computations on $24^3\times 128$ 
and $32^3\times 256$ anisotropic lattices with spatial spacing
$a_s\sim 0.12$~fm and aspect ratio $a_s/a_t\sim 3.5$, where $a_t$ is the
temporal spacing, for pion masses $m_\pi\sim 380$~MeV and $m_\pi\sim 220$~MeV.  
The calculations proceed in several steps: (a) generation of
gauge-field configurations using the Monte Carlo method; (b) computation
of quark sinks for various noises and dilution projectors using the
configurations from the first step; (c) computation of the meson and baryon
sources and sinks using the quark sinks from the second step; (d)
evaluation of the correlators using the hadron sinks; (e) analysis
of the correlators to extract the energies.  Our results for the QCD 
stationary-state energies using, for the first time, both single-hadron
and multi-hadron operators, should appear soon.

This work was supported by the U.S.~National Science Foundation 
under awards PHY-0510020, PHY-0653315, PHY-0704171, PHY-0969863, and 
PHY-0970137 and through TeraGrid resources provided by the Pittsburgh 
Supercomputer Center,  the Texas Advanced Computing Center, and the 
National Institute for Computational Sciences under grant numbers 
TG-PHY100027 and TG-MCA075017.  The USQCD
QDP++/Chroma library\cite{chroma} was used in developing the software
for the calculations reported here.  We thank our colleagues
within the Hadron Spectrum Collaboration.

\bibliographystyle{aipproc}   % if natbib is available

\end{document}